\let\springerBibliographystyle\bibliographystyle
\renewcommand{\bibliographystyle}[1]{\springerBibliographystyle{sn-nature-flovmos}}
\documentclass[pdflatex,sn-nature]{sn-jnl}
\let\bibliographystyle\springerBibliographystyle

\usepackage{amsmath,amssymb}
\usepackage{graphicx}
\usepackage{xcolor}
\usepackage{adjustbox}
\usepackage{booktabs}
\usepackage{multirow}
\usepackage[caption=false]{subfig}
\usepackage{xurl}
\usepackage{lineno}
\usepackage{placeins}

\setcitestyle{numbers,square,comma}
\hypersetup{hidelinks,pdftitle={FloVMos: Optical Flow-based Medical Video Mosaicking}}
\makeatletter
\apptocmd{\ps@headings}{\let\@oddhead\@empty\let\@evenhead\@empty}{}{}
\apptocmd{\ps@titlepage}{\let\@oddhead\@empty\let\@evenhead\@empty}{}{}
\makeatother
\renewcommand{\figurecaptionfont}{\normalfont\fontsize{9}{11}\selectfont\raggedright}

\newcommand{\IV}{\textit{in-vivo }}
\newcommand{\FM}{\textit{FloVMos}~}
\newcommand{\FMs}{\textit{FloVMos}}

\begin{document}

\title{\FMs: Optical Flow-based Medical Video Mosaicking}

\author*[1]{\fnm{Jinyang} \sur{Liu}}\email{liu.jinyan@northeastern.edu}
\author[1]{\fnm{Sandesh} \sur{Ghimire}}\email{drsandeshghimire@gmail.com}
\author[1]{\fnm{Chaman} \sur{Singh}}\email{singh.ch@northeastern.edu}
\author[1]{\fnm{Jennifer} \sur{Dy}}\email{jdy@ece.neu.edu}
\author[2]{\fnm{Milind} \sur{Rajadhyaksha}}\email{RajadhyM@mskcc.org}
\author[1]{\fnm{Dana H.} \sur{Brooks}}\email{brooks@ece.neu.edu}
\author[1]{\fnm{Octavia} \sur{Camps}}\email{camps@ece.neu.edu}
\equalcont{These authors contributed equally to this work.}
\author[2]{\fnm{Kivanc} \sur{Kose}}\email{kosek@mskcc.org}
\equalcont{These authors contributed equally to this work.}

\affil[1]{\orgdiv{Electrical and Computer Engineering}, \orgname{Northeastern University}, \orgaddress{\street{360 Huntington Ave}, \city{Boston}, \postcode{02115}, \state{MA}, \country{USA}}}
\affil[2]{\orgdiv{Dermatology Service}, \orgname{Memorial Sloan Kettering Cancer Center}, \orgaddress{\street{1275 York Ave}, \city{New York}, \postcode{10065}, \state{NY}, \country{USA}}}

\abstract{Biomedical imaging modalities often require a trade-off among resolution, field of view (FOV), and acquisition speed. Video mosaicking offers a strategy to overcome this limitation by computationally stitching sequential high-resolution frames into a wide-FOV composite. However, existing methods struggle with non-rigid deformations, and modality-specific artifacts arising in clinical and research imaging. Here, we present \FMs, a generalizable, optical-flow-based deep learning framework for real-time video mosaicking across diverse biomedical imaging modalities. FloVMos achieves robust, pixel-level registration by fine-tuning an optical flow model on synthetic training data with ground-truth deformation fields. We introduce a pipeline for generating this training data, simulating realistic tissue motion and imaging distortions from existing mosaics or raw videos. Our automated synthetic data generation and optical flow model training based on this data allow users to adapt \FM to different imaging modalities. To demonstrate this, we applied \FM to seven diverse imaging modalities: reflection confocal microscopy, open-top light-sheet microscopy, fetoscopy, laparoscopy, dermoscopy, sparse spectral microscopy, and endoscopy. \FM  outperforms conventional baselines in accuracy, robustness, and speed for all the tested modalities. This adaptable and training-efficient framework enables large-area visualization with real-time performance and may support broader use of video-based biomedical imaging in research and clinical workflows.}


\maketitle
\raggedbottom
\raggedright

\section{Introduction}\label{sec:intro}

There is an intrinsic tradeoff among resolution, field of view (FOV), and imaging speed in many biomedical clinical and research imaging scenarios. This is especially true for modalities such as endoscopy, \IV microscopy, where the imaging device is in contact with, or extremely close to, the surface of the tissue under examination. One attractive solution to this problem is to keep the imaging FOV small, thus allowing both high-resolution and rapid acquisition, and then combine images post hoc into a mosaic to extend the FOV. In effect, then, one acquires a sequence (a ``video") of high-resolution images and then turns them into a single combined image mosaic. Mosaics can reveal spatial relationships in the tissue under examination that occur at a scale larger than the FOV, and thus can provide crucial information to scientists and clinicians about critical structures in the tissue~\cite{yelamos2017margin,bano2020deep,zhou2021real,reeff2006mosaicing}.

Methods to turn videos into mosaics are ubiquitous, including on modern smartphones, and there are a number of useful videomosaicking algorithms described in the literature.  However, biomedical imaging applications often have exigent requirements that cannot be met by existing algorithms. This is particularly true if the path of the imaging device is controlled freehand by the user or is otherwise loosely constrained, as well as if the imaging device can change its angle with respect to the tissue and/or is in direct contact with it. These features cause frame-to-frame changes and tissue distortions that are not accounted for in generic videomosaicking. The few mosaicking methods that have been developed to handle non-rigid deformations are specialized to specific imaging modalities and acquisition pathways and are computationally costly~\cite{zhou2021real,miettinen2019nrstitcher}, resulting in both limited operator control and lack of real-time performance.

Thus, the development of high-resolution, large FOV imaging is impeded by the lack of a general solution to the videomosaicking problem. What is needed is an approach to videomosaicking that is easy to use, fast, and readily adaptable to multiple imaging modalities, including novel modalities that are likely to be developed in the future. The availability of such a solution would provide a dramatic boost both for existing modalities and for future developments.

\begin{figure}[!htbp]
    \centering
    \begin{adjustbox}{max width=\linewidth,max totalheight=0.26\textheight}
    \begin{minipage}{\linewidth}
    \centering
\includegraphics[width=0.98\textwidth]{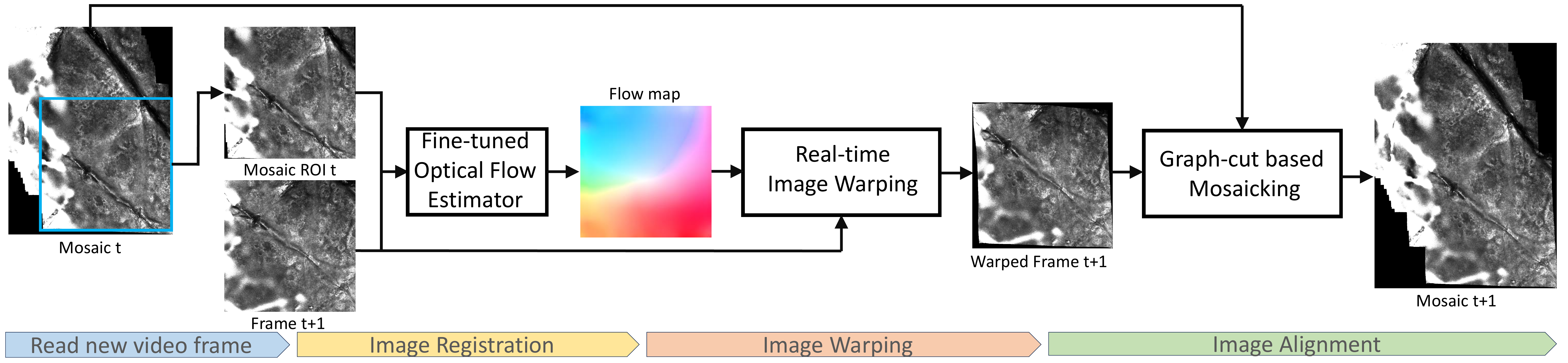}
    \end{minipage}
    \end{adjustbox}
    \par
    \caption{An overview of the \FM videomosaicking pipeline highlighting its four key steps: (i) data acquisition, (ii) image registration using a fine-tuned optical flow estimator (a deep neural network), (iii) real-time image warping, and (iv) image alignment and blending.}
    \label{fig:concisepipeline}
\end{figure}

In this paper, we describe a new solution for biomedical videomosaicking, \textbf{\FMs}, which can rapidly create composite image mosaics to allow seamless real-time imaging of large tissue areas in a variety of imaging applications. %
An overview of  \FM  is shown in Fig.~\ref{fig:concisepipeline}. The heart of \FM is a deep neural network (DNN)~\cite{lecun2015deep} for optical flow-based~\cite{gibson1950perception,gibson1966senses,szeliski2022computer} image registration that employs a supervised training scheme. Using a DNN allows \FM both to handle non-rigid registration of deformed image pairs and to adapt to a wide variety of imaging modalities by simply retraining the underlying DNN with samples from a new modality. The four main steps of \FM are data acquisition, application of the optical flow DNN to register each frame to a growing mosaic, warping of that frame to the mosaic, and application of a graph-cuts-based blending algorithm \cite{Kose_Scientific} designed to minimize image distortions. Each image, or frame, is registered and blended into a growing mosaic as it is acquired, forestalling potential distortions due to error accumulation, a common issue with simpler frame-to-frame registration methods. To preserve processing speed, \FM tracks a moving region of interest on the mosaic and registers and blends each new frame to this region of interest.

To facilitate the adaptation of \FM across modalities, we developed a pair of novel strategies to synthetically generate the data with known optical flow parameters that are required to train the DNN, either from existing mosaics or from videos, both without manual human labeling (details in  Sec.~\ref{sec:methods}). These synthetically generated datasets are used to fine-tune the parameters of the optical flow estimation DNN, ensuring image registration that is tailored for a particular imaging modality. Once the optical flow model is trained, application to new sequences occurs in an ``inference" mode (Fig.~\ref{fig:concisepipeline}), which enables computationally cheap rapid processing. \FM can perform videomosaicking of 1 MP images at ten frames per second (fps), even in our research code implementation, fast enough to be considered real-time for many \IV medical imaging modalities such as microscopy and endoscopy. Finally, \FM also includes an outlier frame detection and buffering strategy to handle abrupt motion (``jumps") in the videos and maintain mosaicking continuity while minimizing errors.

We report on tests of \FM on image sequences from 7 different modalities: reflection confocal microscopy (RCM) of skin tissue, open-top light-sheet microscopy (OTLS), fetoscopy, laparoscopy, dermoscopy, sparse spectral microscopy (SSM), and endoscopy. In each case, we were able to rapidly retrain the DNN and then apply it successfully to the new modality. Our data generation, videomosaicking codes, and optical flow estimation models are available for public access at
\url{https://github.com/JinyangMarkLiu/OpticalFlowBasedMosaicking}.

\FloatBarrier
\section{Methods}
\label{sec:methods}

This section details our \FM implementation. We first outline the \FM framework in Sec.~\ref{sec:briefdes} and then describe our novel synthetic data-based training method and each framework component in Sections~\ref{sec: TrainingData}-
\ref{sec:WarpAndStitch}. These sections are followed by our approach to handle video stream jumps (Sec.~\ref{subsec:outlier}) and our salient-points-based method for assessing mosaicking accuracy without ground truth (Sec.~\ref{sec:SalPts}). Finally, Sec.~\ref{sec:Datasets} details the data used in our experiments.

\subsection{Brief Description of \FM operation}
\label{sec:briefdes}

\begin{figure}[!htbp]
    \centering
    \begin{adjustbox}{max width=\linewidth,max totalheight=0.40\textheight}
    \begin{minipage}{\linewidth}
    \centering
\includegraphics[width=0.85\textwidth,height=0.6\textheight,keepaspectratio]{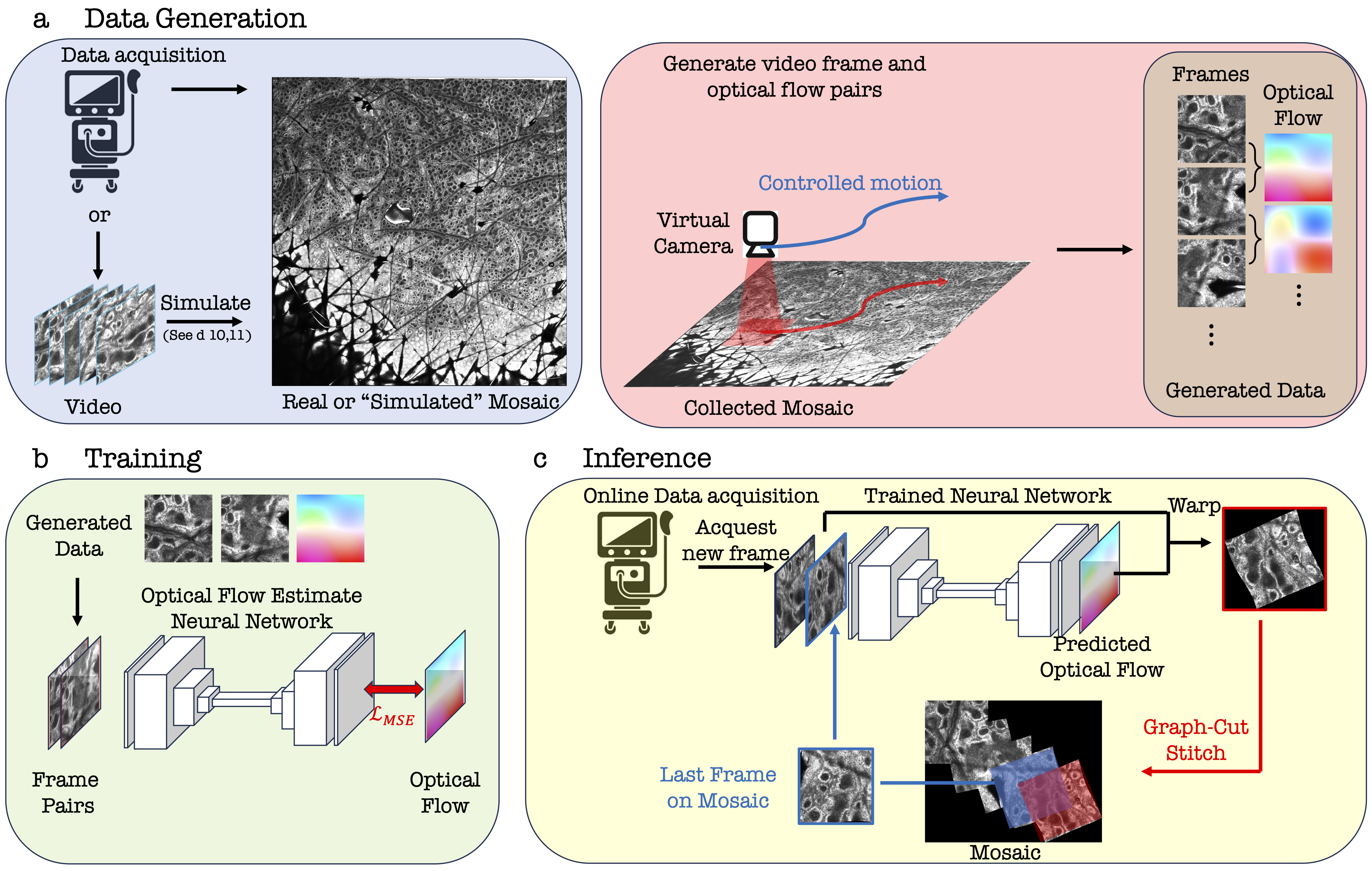}
    \end{minipage}
    \end{adjustbox}
    \par
    \caption{
    Overview of \FMs: \textbf{a} The first step in training involves the acquisition of training data. We have developed two approaches for easy, rapid, and automated acquisition of the required training data. In the first, when sample mosaics are easily available, we simply simulate videos by following a randomized but known path through the mosaic and then adding noise and distortion, as illustrated in the figure. In the second, when we only have videos available, we randomly choose or simulate pairs of images from the video, again with known relationships between the elements of each pair, and add noise and distortion. In both cases, we can thus assemble a collection of labeled pairs of images to use as training data.
    \textbf{b} We fine-tune the optical flow neural network using the annotated training data.
    \textbf{c}
    After training, we apply the DNN to newly acquired (or ``test") videos. The video frames are registered using DNN-predicted optical flow
     and then assembled into a mosaic by our graph-cuts based stitching algorithm.  appending each registered image to the growing mosaic. }
    \label{fig:pipeline}
\end{figure}

\FM is illustrated in Fig.~\ref{fig:pipeline}. As shown in the figure and described in the caption, \FM first undergoes a training process where the parameters of the optical flow DNN are fine-tuned to a particular modality. This training uses pairs of images from the modality of interest, which are related to each other by known optical flow parameters. Our procedures to automatically extract this training data from existing mosaics or videos from that modality are described in detail in Sec.~\ref{sec:methods}. Once the DNN has been fine-tuned, \FM is ready to process newly acquired videos. First, the optical flow DNN estimates optical flow between each new frame and ``a cropped region-of-interest" covering the most recently acquired frame already integrated into the mosaic. The registered images are then added to the existing mosaic by blending the new frame with the existing mosaic. This process is repeated on each frame until the end of the video. If any abrupt jump (e.g., motion blur, depth changes) is detected, based on an abrupt loss of optical flow smoothness as detailed in Sec.~\ref{subsec:outlier}, we either modify the mosaicking process to overcome the jump or, if that is not feasible, terminate the videomosaicing process, saving the existing mosaic,  and begin a new mosaic starting from that image frame.

\subsection{Automated Generation of Training Data}
\label{sec: TrainingData}

Accurate image registration underpins all video mosaicking techniques, serving as a critical foundation for downstream processing. Optical flow-based registration, which estimates dense, pixel-wise displacement fields, is particularly well-suited to capturing the complex non-rigid deformations frequently observed in medical imaging. However, state-of-the-art optical flow models trained on everyday scenes often produce large errors or fail entirely when applied to domain-specific medical data. To address this gap, we developed an automated training data generation framework that produces synthetic video sequences paired with ground-truth optical flow annotations. This framework simulates realistic camera motion and tissue deformation, enabling effective domain adaptation of optical flow estimators to a wide range of imaging modalities.

In some applications, large field-of-view images may already be available, either through an alternative version of the imaging device or generated using specialized offline programs specifically developed for the target application. For example, in one of our modalities of interest, reflectance confocal microscopy (RCM) imaging of skin, it is possible to collect mosaic images using an articulated arm configuration of the microscope, which latches to the area of interest and moves the objective lens over the imaging area in a controlled fashion (e.g., via a raster or strip scan). In another modality, open-top light-sheet (OTLS) microscopy, data is acquired in large-scale mosaic volumes by mounting multiple optically cleared specimens on a flat transparent holder, with both the illumination and collection objectives positioned beneath the imaging plane. For these modalities, we simulate videos by simulating the path of a limited FOV device across a given larger mosaic, starting at a random location in the interior of the large FOV image and following a randomized path with controlled random parameters. Since the process is randomized, we can repeat it many times to acquire as large a set of simulated videos as desired.

For modalities where only videos are available, in which each individual image (frame) has a limited field of view, we simulated the acquisition of image pairs with known optical flow relationships using individual video frames in two different ways. The first method involves randomly selecting a single frame from a video and tiling this frame into a $2\times 2$ grid of frames by flipping the frame vertically and horizontally, resulting in smooth, synthetic ``mini-mosaics" that are twice the size of the original frames along each dimension.%
The second method involves randomly selecting four frames that are temporally distant (at least 15\% of the video's total length apart) from each other within the video sequence. We further (manually or automatically) ensure that there is less than 25\% overlapping area between each of these selected frames to prevent duplicating nearby regions, in order to minimize potential confusion during optical flow training. These four frames are then again arranged into a $2\times 2$ grid.%
While these simulated mini-mosaics may often appear unrealistic %
we found that they serve as a practical alternative for training data acquisition when large field-of-view images or mosaics are not available.

These larger field-of-view images, whichever way they are acquired, can then be used to generate multiple synthetic videos with optical flow annotation. The synthetic data generation process starts by randomly selecting a location within the mosaic and extracting the initial frame matching the field of view of the target imaging modality. The rectangular area covered by this first frame serves as the initial synthetic camera view. Subsequent frames are generated by first randomly synthesizing an optical flow map that mimics both camera motion and the potential non-rigid deformations commonly observed during data acquisition. Using this optical flow map, we perform backward warping to produce the next frame. This process is repeated---generating a new optical flow map and corresponding frame---until the synthetic camera view reaches the edge of the mosaic. Note that for the first method, where simulated videos are extracted from large FOV mosaics, there will be many frames in the simulated video, while for the second method, where we start with a $2\times 2$ synthetic image, there will be far fewer frames in each video. However, by repeatedly applying both processes to construct these $2 \times 2$ grids, we can still acquire as many image pairs for training as desired.

In more detail, each optical flow map used in either approach is randomly constructed by combining a global affine transformation (represented by a homography~\cite{hartley2003multiple}) with a denser non-rigid deformation field. (All parameters defined below are chosen using the values shown in Table~\ref{tab:training-params}.) The affine component is parameterized by translation, rotation, and shearing. To ensure smooth and realistic camera motion, we adopt an autoregressive motion model, in which the translation and rotation parameters at each step are conditioned on both the values from the previous step and a newly sampled random speed for the current step. The parameters are chosen as follows:
Translation speed parameters are sampled from a uniform distribution within predefined bounds, denoted as \(\text{Min Trans.}\) and \(\text{Max Trans.}\). Rotation speed and shearing are applied with probabilities sampled from uniform distributions in the ranges \([0, \text{Max Rot.}]\) and \([0, \text{Max Shear}]\).

\begin{equation}
    \begin{aligned}
    T_{i} &=& \alpha_1 T_{i-1} + (1-\alpha_1) V_i \quad \quad \quad &\text{Translation} \\
    \theta_{i} &=&  \alpha_2 \theta_{i-1} + (1-\alpha_2) \omega_i \quad \quad \quad & \text{Rotation angle} \\
    \end{aligned}
\end{equation}
where $i$ is the frame number in the synthetic video, and $\alpha_1$, $\alpha_2$ are set to 0.9 and 0.5 separately, defining how the rotation and translation parameters are related to the previous state. Here, \( V_i \) and \( \omega_i \) represent the sampled translation and rotation speeds, while \( T_i \) and \( \theta_i \) denote the resulting translation vector and rotation angle applied to construct the global affine transformation.

In order to simulate non-rigid deformations, we then deform the rigid optical flow field on a frame by frame basis by adding a smooth noise field to it with a preset probability \( P_{\text{non-rigid}} \). To ensure the smoothness of the noise field, we first generate it as a coarse field of size $4\times 4$ where the coefficients are drawn from a uniform distribution within the range $[- \textit{Non-Rigid Level}, \allowbreak \textit{Non-Rigid Level}]$ (see Table~\ref{tab:training-params} for parameter values). We then upsample this coarse field to the size of the camera FOV via bilinear interpolation, ensuring the noise field is smooth.

As the variance of the noise increases, the non-rigid deformations introduced in the generated data also increase. This combination of projective transformation and non-rigid deformation allows us to generate the optical flow field between each pair of consecutive synthetic video frames. By varying the parameters governing the projective transformations and non-rigid deformations during the data generation process, we can tailor the camera motion, speed, and degree of non-rigid distortion to match the characteristics of different imaging modalities. For example, using the degree of non-rigid deformation as an example, we can compare dermoscopy with RCM. In dermoscopy, the imaging process typically involves a stable contact between the device and the skin surface, resulting in relatively minor non-rigid motion artifacts. Therefore, we set the \textit{Non-Rigid Level} to three for this modality. In contrast, reflectance confocal microscopy (RCM) often involves handheld operation and imaging at higher magnification, which can lead to significant non-rigid motion due to subtle tissue deformation or hand-induced jitter. To reflect this greater variability, we set the \textit{Non-Rigid Level} to 10 for RCM. We similarly set other parameters, including translation, rotation, and shearing, based on prior knowledge or empirical observations specific to each modality. Detailed parameter choices for each modality are summarized in Table.~\ref{tab:training-params}.

\begin{tableorg}[!htbp]
\tablebodyfont
\centering
\caption{Parameters used for the training dataset of each modality. Note that translation parameters listed in the table are defined based on images resized to a resolution of
256\ensuremath{\times}256}
\label{tab:training-params}
\begin{adjustbox}{width=\linewidth}

\begin{tabular}{lcccccccc}
\toprule
\textbf{Modality} & \textbf{Max Trans.} & \textbf{Min Trans.} & \textbf{Max Rot.} & \textbf{Max Shear} & \textbf{Non-Rigid Level} & \textbf{\(P_{\text{rot}}\)} & \textbf{\(P_{\text{shear}}\)} & \textbf{\(P_{\text{non-rigid}}\)} \\
\midrule
Fetoscopy   & 32 px & 16 px  & 10\ensuremath{^\circ} & 0.1 & 5.0 & 0.2 & 0. & 0.5 \\
Dermoscopy  & 32 px  & 16 px  & 10\ensuremath{^\circ} & 0.1 & 3.0 & 0.1 & 0.1 & 0.1 \\
RCM         & 64 px  & 32 px  & 15\ensuremath{^\circ} & 0.1 & 10.0 & 0.2 & 0.2 & 0.05 \\
Endoscopy        & 32 px & 16 px  & 10\ensuremath{^\circ} & 0.1 & 10.0 & 0.2 & 0.2 & 0.5 \\
SSM   & 64 px  & 32 px  & 5\ensuremath{^\circ} & 0.1 & 1.0 & 0.1 & 0.1 & 0.05 \\
Laparoscopy & 32 px & 16 px  & 10\ensuremath{^\circ} & 0.1 & 10.0 & 0.2 & 0.2 & 0.5 \\
OTLS (synthetic) & 32 px  & 16 px & 5\ensuremath{^\circ}  & 0.0 & 1.0 & 0.1 & 0.0 & 0.05 \\
\bottomrule
\end{tabular}
\end{adjustbox}
\end{tableorg}

Once we have obtained this simulated noisy training data, we can use it to fine-tune our DNN optical flow estimator networks for any particular modality as described in the next subsection. %

\subsection{Optical Flow DNNs}
\label{sec:OpFlowDNNS}

We use the training data obtained through the automated procedure described in the previous subsection to fine-tune existing optical flow DNNs described in the literature. Our base DNN was \textit{FlowNet2}~\cite{ilg2017flownet}, but as we describe in Sec.~\ref{sec:results_optflowcomparisons}, we also tested with three additional optical flow DNNs (\textit{RAFT}\cite{teed2020raft}, \textit{GMFlow}\cite{xu2022gmflow}, and \textit{FlowFormer}\cite{huang2022flowformer}) to assess how sensitive \FM is to the specific choice of DNN. The base FlowNet2 model weights were obtained from its official PyTorch repository (\url{https://github.com/NVIDIA/flownet2-pytorch}). All four of these models are publicly available with pre-trained coefficients. We used the training data to adapt these pre-trained coefficients for each of the seven imaging modalities we tested, in each case starting from the published model coefficients.

We trained each model on a computer equipped with an NVIDIA RTX 2080 Ti X GPU with 11GB VRAM, AMD Ryzen 3900X, and 32 GB of memory. We used a batch size of 8 for \textit{FlowNet2} and \textit{GMFlow} and 6 for \textit{RAFT} and \textit{FlowFormer}. These batch sizes were the maximum that fit within the 11 GB VRAM limit for each model. We employed the Adam optimizer~\cite{kingma2014adam} with an initial learning rate of $10^{-4}$, for 100 epochs. To speed up the training process, we resized the images to 256 pixels $\times$ 256 pixels for all models. The original resolutions for different modalities are as follows: dermoscopy, RCM, SSM, and OTLS each have a resolution of 1000\(\times\)1000 pixels; endoscopy videos are typically acquired at a resolution of \(1280 \times 1024\) pixels; the resolutions for fetoscopy and laparoscopy vary depending on the specific implementation: fetoscopy videos range from \(380 \times 380\) to \(720 \times 720\) pixels, while laparoscopy videos range from \(320 \times 240\) to \(720 \times 288\) pixels.

The result of this fine-tuning is a modality-adapted DNN optical flow estimator that we then applied to incoming image pairs to estimate the optical flow transformation for image-to-mosaic registration, as described next.

\subsection{Image to Mosaic Registration}
\label{sec:ImageToMosaic}

Once we have estimated optical flow parameters between all pairs of consecutive images in a video, we could have taken a naive approach and simply register all images to the first frame. However, this method requires us to update the optical flow field recursively, and as we progress through the video, small errors at each step of the optical flow estimation process can accumulate and even potentially lead to a complete stitching failure \cite{Kose_Scientific}. To mitigate such an accumulation of registration errors, we instead registered each frame directly into the growing mosaic. In other words, during the mosaic-building process, instead of using consecutive video frames to estimate the registration parameters, we apply the DNN to estimate optical flow between the incoming video frame and a cropped, rotated rectangular region from the latest version of the mosaic that corresponds to the most recently stitched frame. Specifically, we track the center and corners of the last stitched frame within the mosaic to estimate its translation and rotation relative to the mosaic coordinate system. Using these transformations, we crop the corresponding ROI from the mosaic and compute the optical flow between the new frame and this reference region.  Once we have the requisite optical flow parameters, we apply inverse translation and rotation to map the registered frame back to the mosaic coordinate system. This means that the pipeline depicted in the ``inference" panel in Fig.~\ref{fig:pipeline} is traversed in its entirety for every frame of the video, but because of our fast, GPU-assisted implementation, we are able to maintain mosaicking speed, as reported in Fig.~\ref{fig:Quantitative}(c).

\subsection{Image Warping and Graph-cut Based Stitching}
\label{sec:WarpAndStitch}
The pixel displacements generated by the optical flow model result in non-integer values for the pixel locations before warping. To resolve this issue, we need to interpolate to an integer-based grid using a scheme designed for non-rigid deformations. Thus, we adopted a bilinear interpolation scheme from the deformable convolution networks literature~\cite{dai2017deformable}. This approach allows us to handle the complexities of non-rigid deformations in the dense flow field, while maintaining the real-time performance essential for video mosaicking. Bilinear interpolation enables smooth sampling and interpolation of intensities from the new frame, ensuring accurate reconstruction of warped frame pixels onto the integer grid.

Specifically, the value \( x(p) \) at the desired integer position \( p \) can be calculated as the convolution with an interpolating kernel:
\begin{equation}
    x(p) = \sum_q G(q,p) \cdot x(q)
\end{equation}
where \( p \)\ and \( q \) are two-dimensional locations in the plane and \( x(q) \) represents the warped frame with fractional positions \( q \). The bilinear interpolation kernel \( G(q,p) \) is defined as \( G(q,p)=g(q_x,p_x) \cdot g(q_y,p_y) \), where \( g(a,b) = \max(0, 1-|a-b|) \). This formulation allows for real-time implementation of image interpolation, with a processing speed of approximately 35 frames per second (in a Python implementation running on the CPU of a regular workstation) on videos with a resolution of 1000\ensuremath{\times}1000 pixels.

Once the frame is registered to the mosaic ROI, we apply a graph-cut-based stitching algorithm we have previously developed in \cite{Kose_Scientific}, to merge the new frame with the mosaic. Briefly, this technique minimizes visual artifacts by stitching the mosaic and the new frame along an optimal seam, which is designed to pass through less detailed (low-pass signal) regions of the overlap between the mosaic and the new frame. The method retains the pixel values of the registered frame and thus avoids any blurring that might occur if we were instead to blend multiple pixel values. However, because the technique identifies a stitching seam between the mosaic and frames, the same pixel in the mosaic could be overwritten multiple times if it falls within the overlap region across more than two frames. In such cases, the final value of the pixel is determined during the graph-cut-based stitching process by whether it is on the mosaic or the latest relevant frame side of the seam.

\subsection{Handling abrupt changes in the videos}
\label{subsec:outlier}
Abrupt motion of the imaging device can result in blurring and/or discontinuities in the video data. Moreover, for 3D imaging modalities such as RCM imaging, abrupt changes in imaging depth can also lead to discontinuities between consecutive frames. These phenomena can lead to severe errors in stitching and ultimately compromise the entire mosaicking process. Our approach is to first detect these abrupt changes and then either use adjacent frames to maintain continuity if possible, or start a new mosaic if that is not possible.

To accomplish this, we developed an efficient method to detect outlier frames using optical flow. We start by observing that, in the absence of abrupt changes, the warping field between consecutive images is non-rigid but smooth. However, when artifacts are present in image data, the optical flow estimator can mistakenly align non-corresponding, yet visually similar, features as it aims to minimize differences between two input images. To detect these irregularities, we employ a simple method that measures the smoothness of the optical flow by averaging a discrete estimate of the spatial second-order derivatives of the flow map. This smoothness metric helps us detect when the neural network's outputs contain artifactual areas. Through empirical analysis---specifically by introducing artifacts until the optical flow network failed to produce a smooth flow---we determined a threshold value of 0.1 for flagging anomalous frames. Once a frame with artifactual areas is flagged, our pipeline first attempts to ensure the continuity of stitching by utilizing the optical flows from subsequent frames following the one that led to distorted flow. If the optical flow generated by a subsequent frame is artifact-free, as determined by this threshold, then the process continues from that frame, discarding the artifact-corrupted frame. Specifically, to maintain continuity in video stitching, when abnormal optical flow is detected, we buffer a number of subsequent frames proportional to the video acquisition frame rate (e.g., buffering 1 second's worth of frames). For each buffered frame, we compute its optical flow relative to the mosaic reference area and evaluate the smoothness of the flow field by calculating its mean second-order derivatives. The frame associated with the smoothest flow (i.e., lowest mean second-order derivatives) is then selected as the next valid frame for stitching, as it is likely to be the first normal frame following the artifact-corrupted frames. However, to ensure the reference frame and the frame selected from the buffer contain enough overlap (since they are not consecutive images in the video), the average pixel value difference between the mosaic region of interest and the warped frame is computed. If this difference exceeds an empirical threshold of 30 (for monochrome images) or 40 (for RGB images), instead of trying to combine that frame with the existing mosaic, we initiate a new mosaic starting from the last buffered frame.

 \subsection{Identifying Salient Points for Computing Registration Errors}\label{sec:SalPts}

In the absence of the ground truth optical flows, it is not possible to quantitatively measure the accuracy of the registration. To address this problem, we developed a novel procedure to estimate registration accuracy. Our primary assumption is that the distance between structures in the frames should not change drastically in the mosaic after registration and warping. In this respect, we developed a method that compares the distance between matching salient points in the input videos and the reconstructed mosaics. Specifically, we extracted salient features from each video frame as well as the mosaic using the Scale-Invariant Feature Transform (SIFT)~\cite{lowe2004distinctive}. To establish correspondences, we applied a brute-force matcher to identify candidate matches between the SIFT keypoints of each frame and those of the mosaic. Given the characteristics of these images, including repeated or structurally similar patterns, erroneous matches are likely to occur. To ensure robustness, we employed the Random Sample Consensus (RANSAC) algorithm~\cite{ransac} with 10,000 iterations to estimate a homography and identify inlier correspondences. Frames with fewer than 5\% inliers, relative to the total number of detected features, were excluded from further analysis. Additionally, we constrained the estimated scaling factors of the homography---defined as the eigenvalues of the \(2 \times 2\) top-left submatrix of the homography matrix---to lie within the range \([0.8, 1.25]\). Matches that produced scaling factors outside this range were considered outliers and subsequently discarded. After this verification of geometric consistency, we computed a pairwise distance matrix between the matched keypoints in the frame and the mosaic. The frame-wise displacement of salient features was then obtained by subtracting the distance matrix of the frame from that of the mosaic. This procedure was then repeated for every frame in the sequence to compute an averaged error metric.

\subsection{Dataset Details}\label{sec:Datasets}

We tested \FM using data from seven different imaging modalities. All RCM and dermoscopy images were collected at the Memorial Sloan Kettering Cancer Center (MSKCC). The OTLS microscopy mosaics and SSM videos were graciously provided by Northeastern University's Spring Lab (\url{https://sites.google.com/view/springlabnu}) and the University of Washington's Molecular Biophotonics Laboratory (\url{https://washington-seattle.digication.com/jonliu/Home}), respectively. The endoscopy, laparoscopy, and fetoscopy videos were obtained from publicly available datasets (the Endoscopic Vision Challenge (\url{https://opencas.dkfz.de/endovis/challenges/2024/}), Hamlyn Center Laparoscopic/Endoscopic Video (\url{https://hamlyn.doc.ic.ac.uk/vision/}), and FetReg datasets (\url{https://www.ucl.ac.uk/interventional-surgical-sciences/weiss-open-research/weiss-open-data-server/fetreg-largescale-multi-centre-fetoscopy-placenta-dataset}), respectively).

For the RCM and OTLS imaging modalities, we had large field-of-view mosaic images available that were generated using a non-handheld version of the image acquisition devices. These devices are capable of moving the sample with respect to the imaging device in a controlled fashion (e.g., raster scan, or strip scan \cite{SanjeeStrip}), using additional hardware (e.g., a motorized stage), and generating mosaics by tiling the collected images. We used these mosaics to generate simulated mosaics as described above and used them either to obtain training data or for testing of our algorithm. Additionally, we had RCM videos that were recorded using a handheld device. These videos are used for testing the framework. For all the other experiments, we only had access to videos. Details about which dataset was used for training and testing are given in Table~\ref{table:dataset}.

\begin{tableorg}[!htbp]
\tablebodyfont
\centering
\caption{Details about each of the datasets used in the work described here. The first column under ``Training Dataset'' indicates if we had access to real mosaics from which we simulated videos for training (indicated by a check mark) or only access to videos (indicated by a cross). The next column shows the number of videos and frames either simulated or available for each training dataset. The first columns under ``Testing Dataset'' indicate whether a given test was carried out using real or simulated videos. Note that for the RCM data we had both available, and in addition (because we had ground truth mosaics for these data) we used separate sets of the simulated videos for different quantitative tests, and report the numbers of videos and frames for each of these testing datasets. The last column indicates the purpose for which we used each of the datasets. For video mosaicking experiments, we excluded sequences with extreme artifacts or strong vertical camera motion that cannot be represented by a 2D mosaic. Furthermore, the same videos were never reused for both generating synthetic training data and testing. $\dagger$: Training set synthesized from real mosaic. $\ddagger$: Testing dataset consists of real videos.}

\begin{adjustbox}{width=\linewidth}
\begin{tabular}{c| c c c| c c c|c}
 Imaging & \multicolumn{3}{c|}{Training Dataset} & \multicolumn{3}{c|}{Testing Dataset}&\multirow{2}{*}{Experiment Type} \\

modality & $\text{Real}^{\dagger}$ &Videos & Frames & $\text{Real}^{\ddagger}$ & Videos & Frames &  \\
\hline
Fetoscopy & \sffamily X & 250 & 500 & \checkmark & 2 & 181 & Video Mosaicking\\
Dermoscopy & \sffamily X& 250 & 500 & \checkmark & 10 & 3,307 & Video Mosaicking \\
\hline
\multirow{4}{*}{RCM}   &\multirow{4}{*}{\checkmark} & \multirow{4}{*}{11,000}  & \multirow{4}{*}{40,000}  & \checkmark & 5 & 1,182 & Video Mosaicking \\
 & & & & \sffamily X & 800 & 17,000 & Speed \& Accuracy Test \\
  & & & & \sffamily X & 1,600 & 3,200 & Handling Non-rigid  \\
   & & & & \sffamily X & 172 & 4,000 & OF Algorithm Sensitivity  \\
\hline
  Endoscopy & \sffamily X& 250 & 500 & \checkmark & 1 & 350 & Video Mosaicking \\
  SSM & \sffamily X& 60 & 120 & \checkmark & 2 & 64 & Video Mosaicking \\
  Laparoscopy& \sffamily X & 250 & 500 & \checkmark & 1 & 428& Video Mosaicking \\
  OTLS & \checkmark& 140 & 2,250 & \sffamily X & 10 & 1,599 & Video Mosaicking\\
\end{tabular}
\end{adjustbox}
\label{table:dataset}
\end{tableorg}

To generate data for the experiment in which we systematically increased the degree of non-rigid deformation, we synthesized videos in which we set the probability of non-rigid deformation ($P_{\text{non-rigid}}$) to 1  and incrementally increased the \textit{Non-Rigid Level} from 0 to 15 in integer steps. For each \textit{Non-Rigid Level}, we generated 100 videos, resulting in a total of 1,600 synthetic videos.

\FloatBarrier
\section{Results}

We present results that illustrate \FMs' ability to easily and accurately mosaic data from diverse imaging modalities.
We then report on three additional technical aspects of \FMs' performance: its ability to handle non-rigid deformations, its mosaicking speed as a function of video sequence length, and its flexibility with regard to the specific optical flow DNN employed.

\subsection{Applying \FM to videos from multiple imaging modalities}\label{sec:Multimodality}

We show results from the application of \FM on eight examples from seven different imaging modalities: fetoscopy,  dermoscopy, reflectance confocal microscopy (RCM), endoscopy,  sparse-spectral microendoscopy (SSM), laparoscopy, and open-top light sheet microscopy (OTLS). For RCM, videos are obtained using a handheld microscope, while mosaics are acquired directly with a different version of the microscope. For OTLS, the mosaics are captured by a bench-top scope. For other modalities, only videos without mosaics are available. Full details on the eight datasets are described in Sec.~\ref{sec:Datasets} and Table~\ref{table:dataset}.

\begin{figure}[!htbp]
    \centering
    \begin{adjustbox}{max width=\linewidth,max totalheight=0.58\textheight}
    \begin{minipage}{\linewidth}
    \centering
\begin{minipage}[b]{\textwidth}
    \centering
    \includegraphics[width=\linewidth,height=0.48\textheight,keepaspectratio]{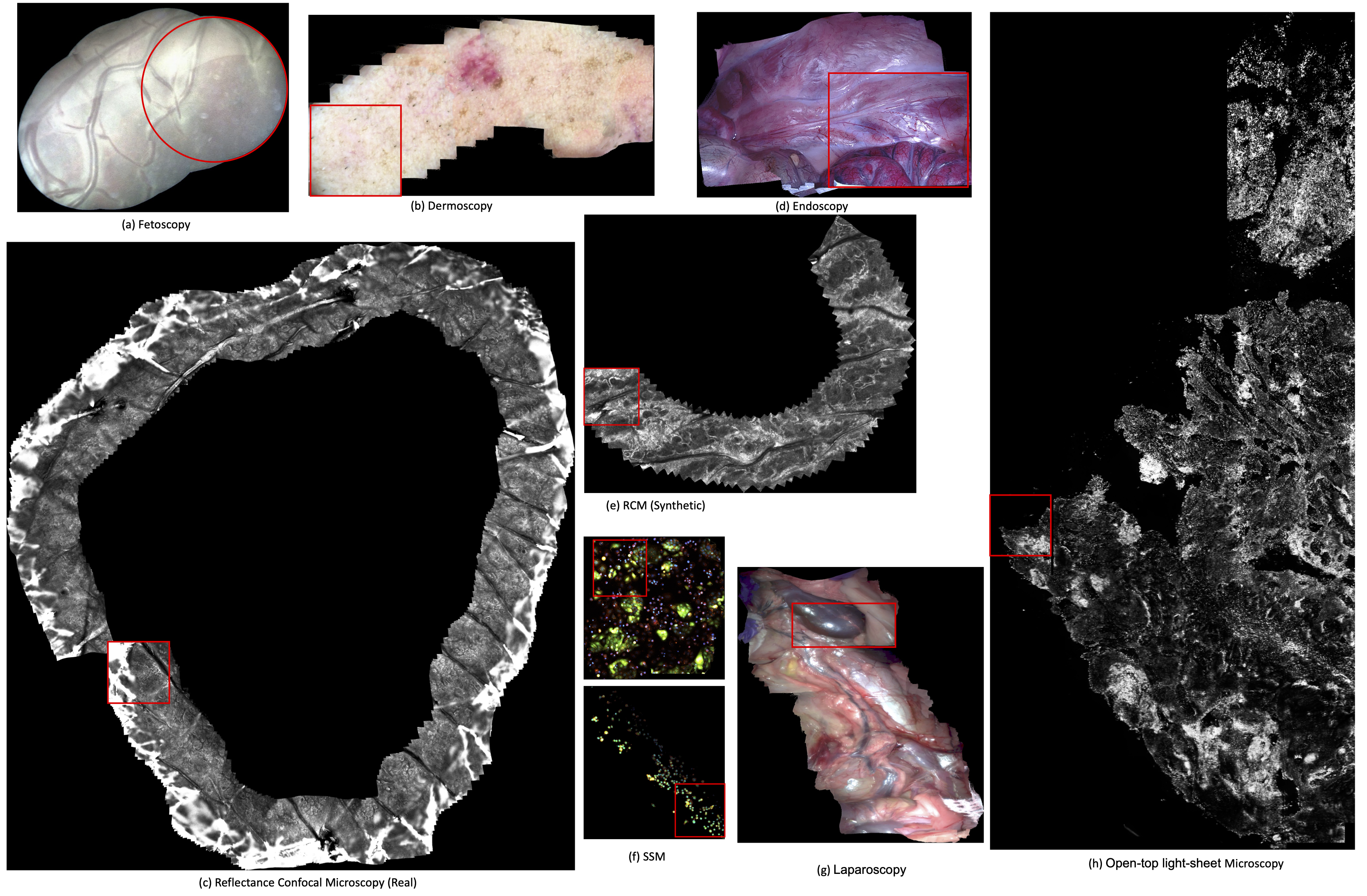}

    \end{minipage}

    \begin{minipage}[b]{\textwidth}
        \centering
        \includegraphics[width=0.8\linewidth,height=0.23\textheight,keepaspectratio]{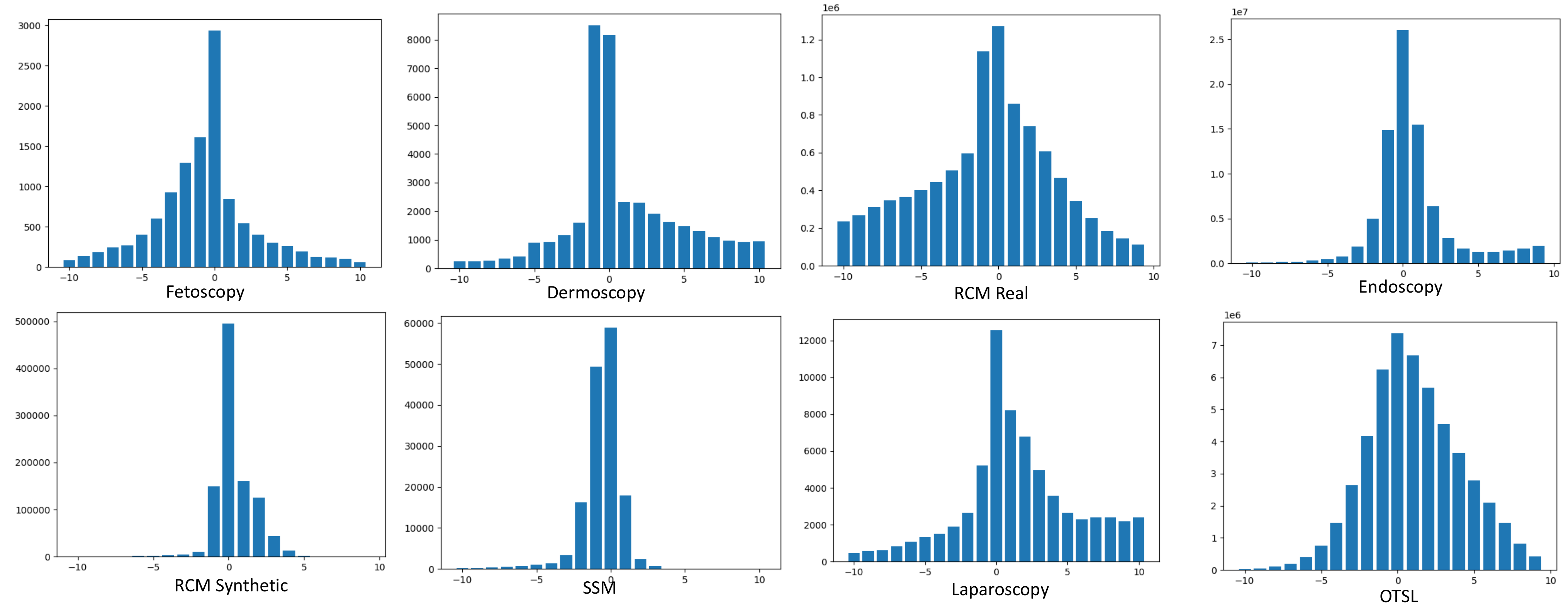}

    \end{minipage}
    \end{minipage}
    \end{adjustbox}
    \par
    \caption{Results of applying \FM to different imaging modalities:  \textbf{Top:} \FM qualitative results are presented for eight cases from seven different imaging modalities: (a) Fetoscopy; (b) Dermoscopy; (c) real RCM; (d) SSM; (e) simulated RCM; (f) Endoscopy; (g) Laparoscopy) and (h) videos simulated   from an OTLS mosaic. The red circle / box in each image shows the location of  the first frame in the video. (We use a red circle for the fetoscopy video to match the actual FOV of the device; when we use \FM to create a mosaic, the frames are stitched onto a black background.) \textbf{Bottom:} Histograms of distances (in pixels) between image feature point pairs in the source mosaics or video frames and their matches in the resulting \FM mosaic.}
    \label{fig:Qualitative}
\end{figure}

We illustrate the operation of \FM on all these datasets in Fig.~\ref{fig:Qualitative} panels \textbf{a} through \textbf{h}. We also report quantitative results on all eight datasets in the bottom panel of Fig.~\ref{fig:Qualitative}.

To calculate quantitative results, we needed some form of ``ground truth" accuracy. For the RCM and OTLS modalities, where we synthesized videos from available mosaics, we simply calculated pixel distances between matched salient point locations in the original mosaic and the reconstructed \FM mosaic. For the other modalities, where we did not have ground truth mosaics, we automatically identified matching salient points in the original video and the reconstructed mosaic and again calculated pixel distances for each case. (Details in Sec.~\ref{sec:SalPts}). Histograms of the resulting pixel distances in the figure are shown for all seven modalities at the bottom of Fig.~\ref{fig:Qualitative}. In the case of RCM we have both real videos and videos simulated from existing mosaics, so we show results for both cases in the figure. In all seven modalities, \FM preserves the distance between feature points after registration, achieving an average of $<\pm5$ pixel/match ($<$0.5\% of the field of view) difference between the feature point pairs in the originals and the matching pairs in the \FM mosaic.

\subsection{Comparison of \FM Videomosaicking Accuracy to Existing Videomosaicking algorithms}

We evaluated \FMs' performance against three state-of-the-art videomosaicking algorithms: Parallax~\cite{zhang2014parallax}, As-Projective-As-Possible (APAP)\cite{Lin_2015_CVPR}, and Automated Video-Mosaicking (AVM)\cite{Kose_Scientific}. We chose RCM as the modality for these comparisons, given our ability to generate ground truth data (simulated videos and corresponding optical flows) that allows us to accurately measure performance under various conditions, including variable non-rigid deformations and different lengths of simulated videos.

\begin{figure}[!htbp]
    \centering
    \begin{adjustbox}{max width=\linewidth,max totalheight=0.57\textheight}
    \begin{minipage}{\linewidth}
    \centering
\subfloat[Quantitative results on the synthetic RCM dataset\label{subfig:performance}]{%
        \begin{minipage}{0.98\textwidth}
            \centering
            \includegraphics[width=0.32\linewidth]{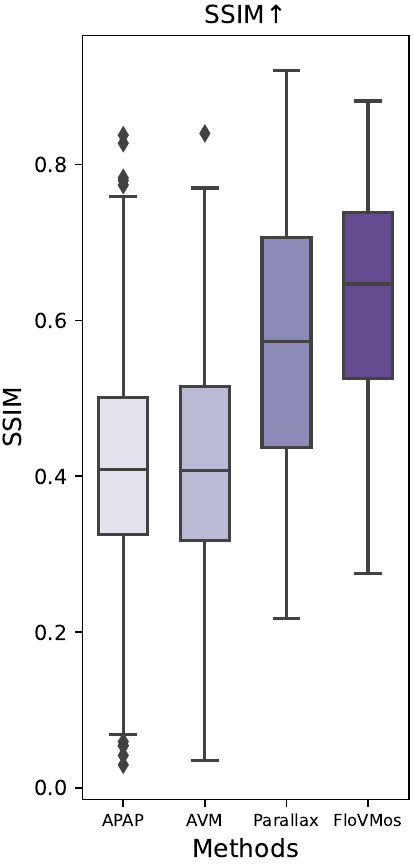}\hfill
            \includegraphics[width=0.32\linewidth]{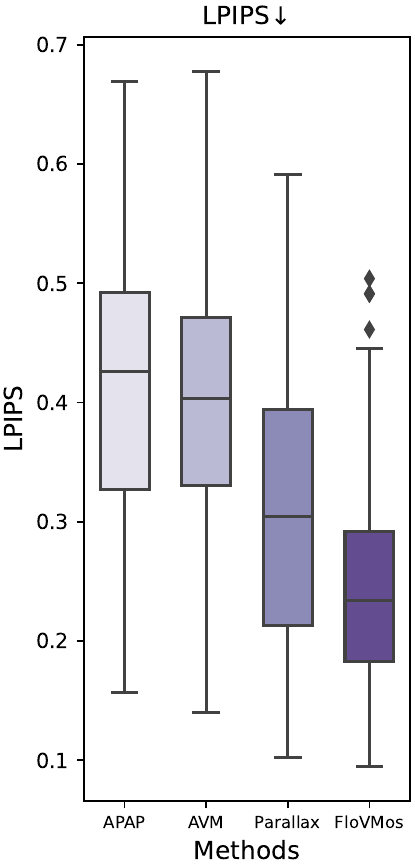}\hfill
            \includegraphics[width=0.32\linewidth]{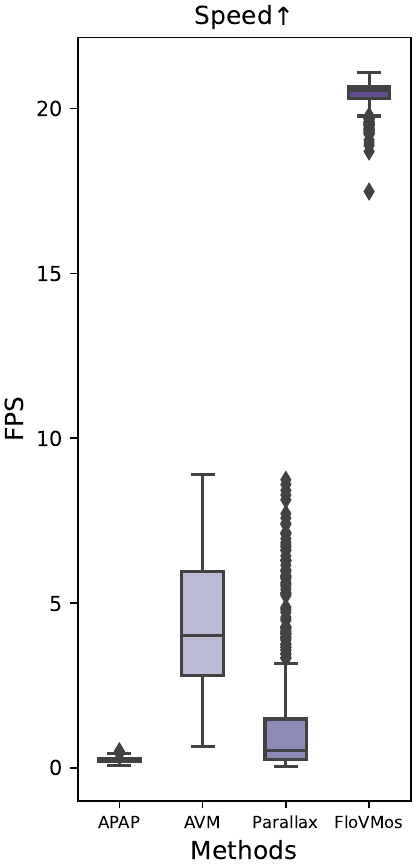}
        \end{minipage}}

    \medskip
    \subfloat[Registration error under non-rigid deformation\label{subfig:regi_error}]{%
        \includegraphics[width=0.48\textwidth]{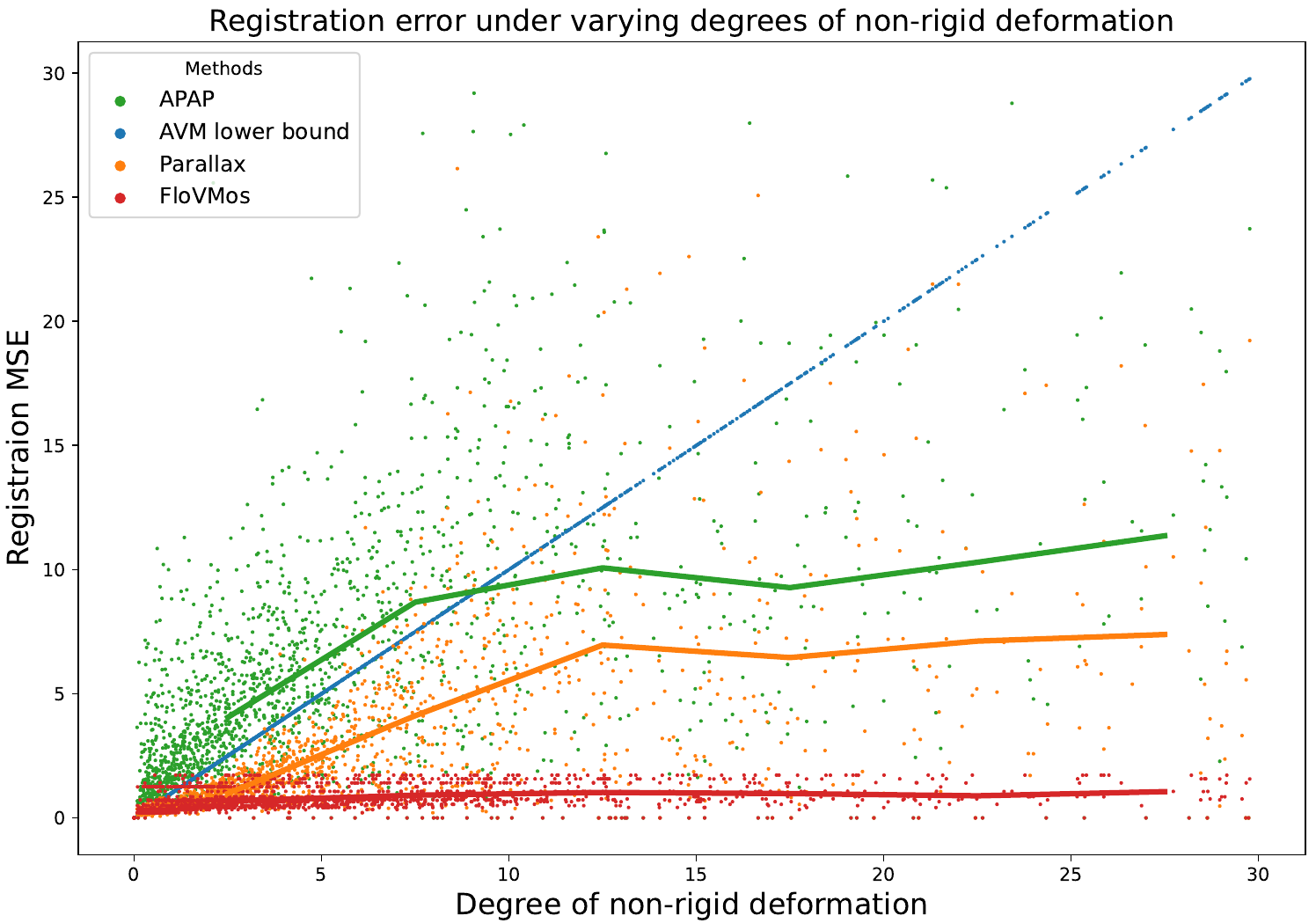}}\hfill
    \subfloat[Speed as a function of video length\label{subfig:speed}]{%
        \includegraphics[width=0.48\textwidth]{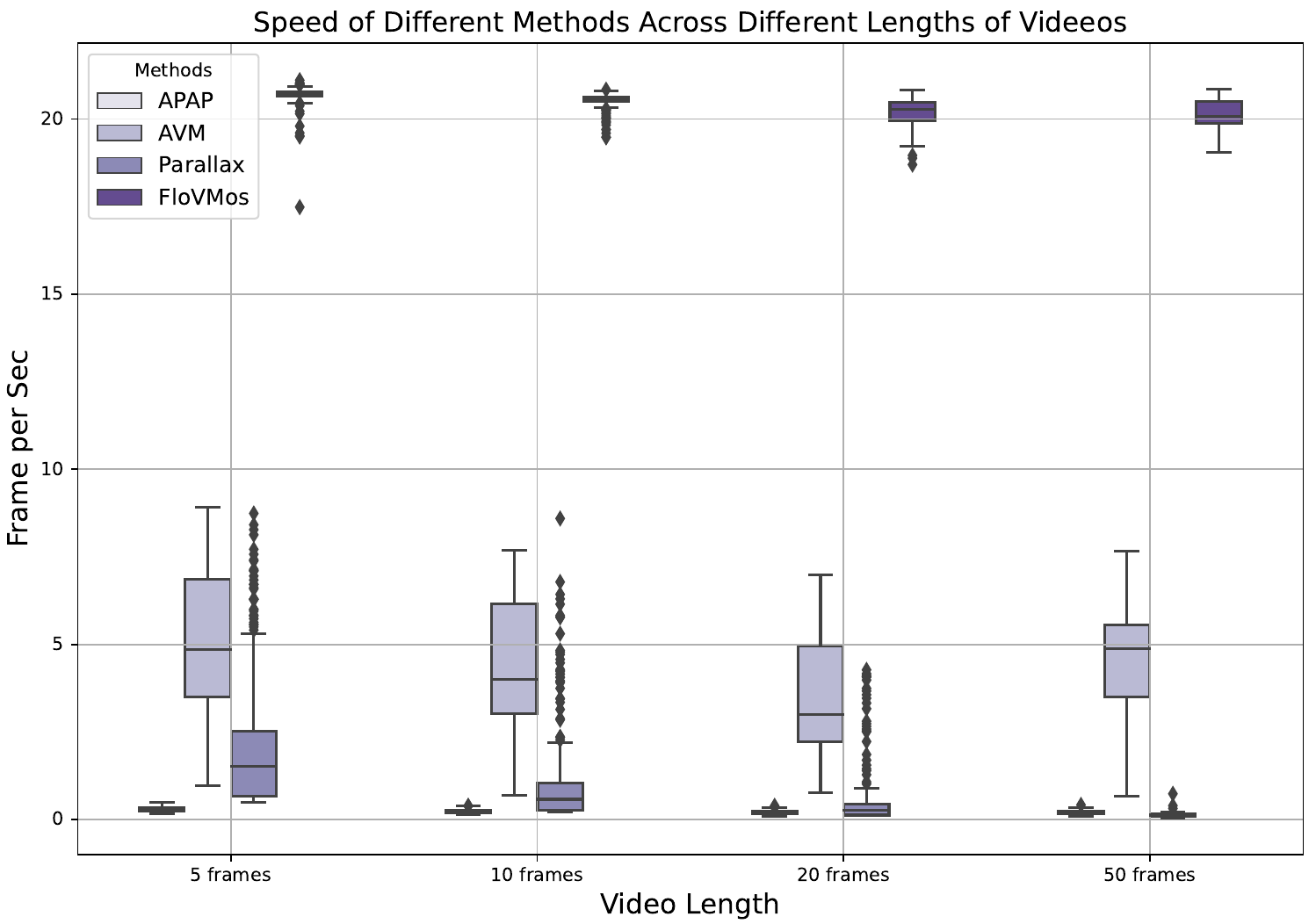}}
    \end{minipage}
    \end{adjustbox}
    \par
    \caption{Quantitative assessment of accuracy and speed of \FM in comparison to selected existing videomosaicking algorithms on the simulated RCM videos:
    (a) Comparison of accuracy of \FM and three comparison algorithms: we compare accuracy using both the Structural Similarity  Index Metric (SSIM, left panel) and Learned Perceptual Image Patch Similarity (LPIPS, middle panel) of \FM to the Parallax,  As-Projective-As-Possible (APAP)~\cite{Lin_2015_CVPR}, and Automated Video-Mosaicking (AVM) algorithms. Note that for SSIM, higher is better, while for LPIPS, lower is better, as indicated by the up and down arrows in the image titles. The right panel reports on mosaicking speed in frames per second (FPS) for the same four algorithms. We show box plots for the full results with  800 simulated RCM videos. The horizontal line inside the box represents the median (50th percentile) of the data. The edges of the box denote the first quartile ( 25th percentile) and the third quartile (75th percentile). (b)  Comparison among the same four algorithms in terms of registration error as a function of increased degree of non-rigid deformation. Colors represent the different algorithms, as indicated in the legend. Each simulated image pair is shown as a dot on the plot. The degree of (artificially induced) non-rigid registration is quantified as explained in the text. (c) Comparison among the same
    four algorithms in terms of mosaicking speed in frames per second. Results are shown in frames per second for videos consisting of 5, 10, 20, and 50 frames, as shown on the horizontal axis. The different algorithms are shown by the color of the box, as indicated in the legend. Again results are shown as box plots for all the synthesized RCM videos, using the same box plot format as in the top panel.
    }
    \label{fig:Quantitative}
\end{figure}

We present our comparative analysis results in Fig.~\ref{fig:Quantitative}(a). We evaluated the accuracy of the videomosaic generated by each algorithm using two standard metrics, the Structured Similarity Index Metric (\textit{SSIM}~\cite{wang2004image}) and the Learned Perceptual Image Patch Similarity (\textit{LPIPS}~\cite{zhang2018unreasonable}) (Fig.~\ref{fig:Quantitative}(a)). Note that for SSIM, higher values mean greater accuracy and the opposite for LPIPS. The panel shows the full distribution of these metrics for the simulated RCM dataset. We observed that \FM resulted in both a better mean value and a more compact distribution than the comparison algorithms. The performance difference compared to even the best of the competing methods, Parallax, is statistically significant (p-value $<$ 0.0001).

\subsection{Handling Non-rigid Deformations}

Non-rigid deformations between video frames can arise from both device control inconsistencies and actual tissue deformation during contact with the device. We conducted extensive experiments to evaluate \FMs' robustness to such deformations, manipulating the presence and severity of these distortions in simulated reflectance confocal microscopy (RCM) videos, where we could control the presence and strength of such deformations in the process of simulating videos from existing mosaics (see the last part of Sec.~\ref{sec:Datasets} for a detailed description). We systematically increased the degree of non-rigid deformation across 1,600 pairs of frames randomly selected from available RCM mosaics (methodological detail in Sec.~\ref{sec: TrainingData}).

A comparison of registration error (measured as mean squared error (MSE) between ground-truth optical flow and predicted optical flow) for \FM and the three comparison algorithms is shown in Fig.~\ref{fig:Quantitative}(b). Each video is represented by a dot, and the color indicates which algorithm was used to do the videomosaicking. Our results show that \FM consistently achieved lower MSE values at various degrees of deformation compared to the competing methods. %

\subsection{Sensitivity of Videomosaicking Speed to Video Length}

A key objective of our work was to develop an algorithm capable of constructing videomosaics in effective real-time, allowing for immediate presentation of mosaics to the device operator. This capability will enable researchers or clinicians to make rapid decisions about where to acquire further images on the tissue being examined, streamlining the imaging process.

To assess the speed of \FMs, we compared its performance to that of the same three alternative algorithms, measuring reconstructed frames per second as we increased the number of frames in the video. Our results shown in Fig.~\ref{fig:Quantitative}(c) indicate that \FM consistently outperforms the other compared methods in terms of speed.

\subsection{Sensitivity to choice of optical flow estimation algorithm}\label{sec:results_optflowcomparisons}

Given that the field of DNNs is evolving rapidly, with new approaches constantly being developed and introduced, we designed the architecture of \FM so that it is modular with respect to the particular optical flow estimation algorithm used. To test the sensitivity of \FM to different optical flow estimators, we ran \FM utilizing four different optical flow networks on our simulated RCM dataset: the standard algorithm used in \FMs, Flownet2~\cite{ilg2017flownet},  Recurrent All-Pairs Field Transforms for
Optical Flow (RAFT)~\cite{teed2020raft}, A Transformer Architecture for Optical Flow (FlowFormer)~\cite{huang2022flowformer}, and Learning Optical Flow via Global Matching (GMFlow)~\cite{xu2022gmflow}.

We trained (fine-tuned) each of the four networks on the same 11,000 videos with 40,000 frames and evaluated them on a distinct set of 172 videos with 6,000 frames. We emphasize that the rest of the \FM algorithm remained the same in all cases; the only change we made was the optical flow network employed. In Table~\ref{table:regi_error} we report the performance accuracy of four different optical flow networks in terms of MSE (the average $\ell_2$ error over all pixels, lower is better), F1-score, which reports the mean number of pixels with endpoint errors greater than 3 pixels, again lower is better), and videomosaicking speed (in FPS)  using the particular optical flow algorithm.

\begin{tableorg}[!htbp]
\tablebodyfont
\caption{Comparison of \FM performance with four different optical flow estimator algorithms on the same (synthetic RCM) data. Each pair of rows shows results for one algorithm: the upper row of each pair reports results when the algorithm is applied ``off the shelf," while the lower row reports results after it is retrained for the particular modality using training data acquired with our automated scheme. The columns show results for average Mean Squared Error (MSE) and F1 score (in both cases, lower is better), and registration speed in frames per second(FPS)  (higher is better). }
\label{table:regi_error}
\centering
\begin{tabular}{l c c c}
\hline
     Method & MSE \ensuremath{\downarrow} & F1 \ensuremath{\downarrow} & FPS \ensuremath{\uparrow} \\
\hline
    FlowNet2 &0.940 & 5.002 & 32.79 \\
    FlowNet2(fine-tuned)& \textbf{0.046} &0.004& 32.79 \\
    \hline
    RAFT &0.921 &8.349 & 10.76\\
    RAFT (fine-tuned) & 0.102 &\textbf{ 0.001} & 10.76 \\
    \hline
    FlowFormer & 2.503 & 14.06 &  14.45 \\
    FlowFormer (fine-tuned) & 0.063 & \textbf{0.001} & 14.45 \\
    \hline
    GMFlow & 1.259 & 4.950 & 35.08 \\
    GMFlow (fine-tuned) &  0.129 & 0.028 & \textbf{35.08} \\
\hline
\end{tabular}
\end{tableorg}

\FloatBarrier
\section{Discussion}

The \FM framework provides a novel solution for the challenging task of image registration and mosaic construction over large tissue areas with limited FOV medical devices. By leveraging an optical flow-based deep neural network based dense image registration model, \FM effectively addresses the critical challenges for wide and easy applicability of videomosaicking for small FOV medical and biological imaging devices, achieving broad applicability across multiple imaging modalities while requiring minimal human effort to retrain for a given modality, robust registration under varying motion and tissue deformation scenarios, and effectively real-time processing. We emphasize that, aside from tuning a few training parameters, applying \FM is a completely automated process, from generating the training data from either existing mosaics or existing videos all the way through the creation of the reconstructed mosaics.

Our results show that \FM can be readily applied to a variety of modalities: as reported here, we tested it on RCM, dermoscopy, endoscopy, fetoscope, OTLS, SSM, and laparoscopy. We believe that this is a broad enough range of modalities to suggest that \FM will work with other existing, or even not-as-yet-invented, modalities as well.  We report that \FM not only achieved lower MSE values compared to comparison methods but also demonstrated improved robustness against increasing levels of non-rigid deformation. We speculate this is true at least in part because of the flexible non-linear processing in the DNN optical flow estimator. While two of the comparison algorithms, along with \FM, showed little variation in processing time with respect to video length (Note that AVM often collapsed midway through stitching when processing 50-framelength videos. As a result, in practice, it ended up stitching shorter segments, which likely contributed to the observed increase in average speed in Fig.~\ref{fig:Quantitative}), in contrast to the Parallax algorithm that went from 0.2 to 0.05 FPS as the video lengths increased, \FM was not only insensitive to video length but also much faster than the comparison algorithms, which we attribute both to our approach of only operating on the most recently stitched part of the growing mosaic and the highly parallelizable nature of the DNN optical flow model, which can be run on a GPU with very high efficiency. Finally, we saw that \FM is relatively robust to the choice of optical flow estimator and confirmed the efficacy of our fine-tuning retraining for both accuracy metrics and all optical flow estimators.

\begin{figure}[!htbp]
    \centering
    \begin{adjustbox}{max width=\linewidth,max totalheight=0.43\textheight}
    \begin{minipage}{\linewidth}
    \centering
\begin{minipage}[b]{0.9\textwidth}
        \centering
        \includegraphics[width=\linewidth,height=0.65\textheight,keepaspectratio]{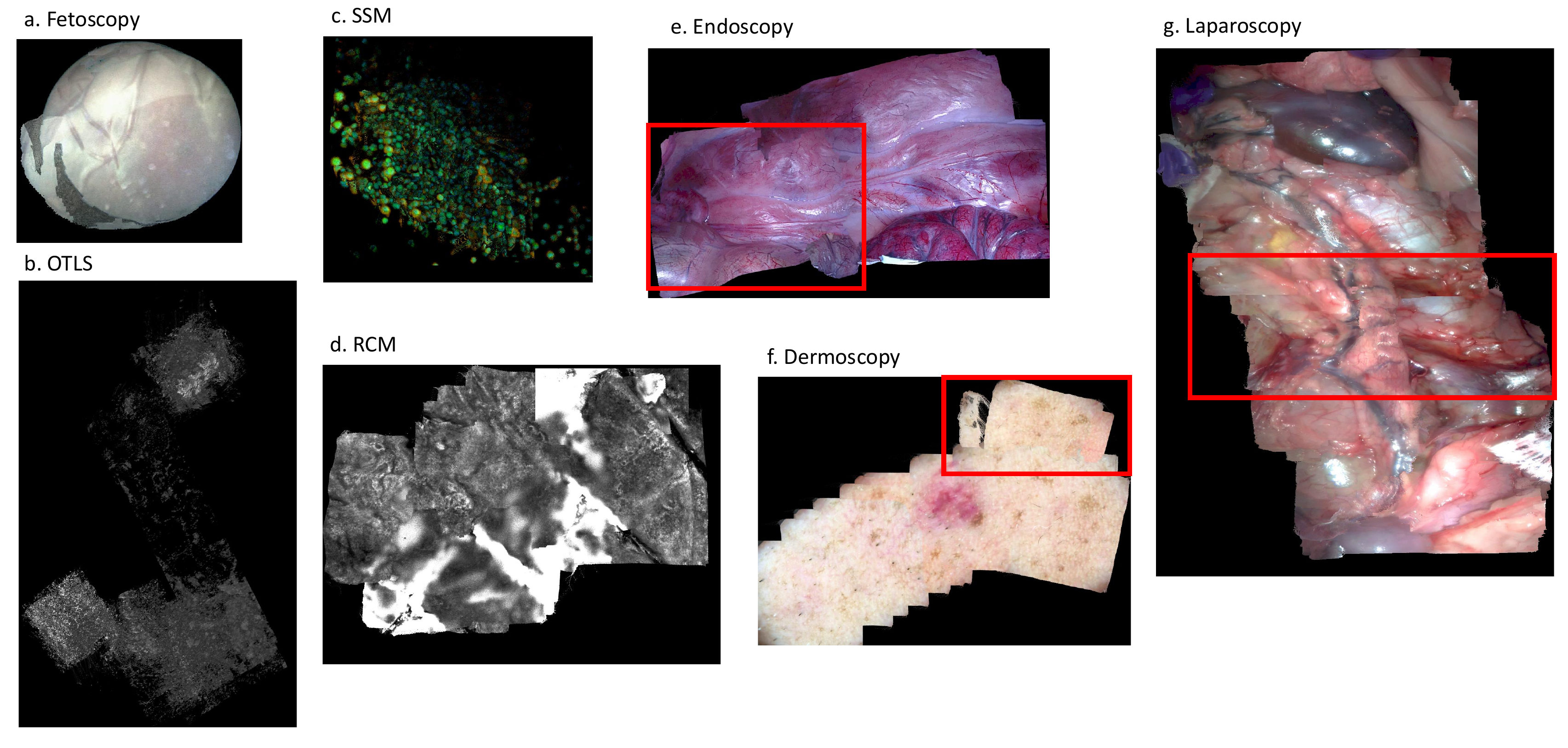}

    \end{minipage}
    \end{minipage}
    \end{adjustbox}
    \par
    \caption{Stitching results obtained using the base optical flow model on the video sequences presented in Fig.~\ref{fig:Qualitative}. The left two columns illustrate modalities characterized by large featureless black or white regions within frames (fetoscopy, OTLS, SSM, and RCM), whereas the right column depicts modalities visually more akin to conventional RGB imagery. Modalities featuring substantial black or white regions often exhibit failures in registration and subsequent stitching. Modalities resembling standard RGB images generally demonstrate improved registration performance even without fine-tuning; however, registration failures still occur. For example, incorrect alignment of the last several frames is evident in (e) endoscopy and (f) dermoscopy, and less accurate optical flow predictions result in noticeable artifacts in the final mosaic, as exemplified in (e) laparoscopy. }
    \label{fig:OffShelf}
\end{figure}

Optical flow-based dense image registration~\cite{gibson1950perception} relies on pixel displacement predictions represented as vector fields. Given an optical flow field, in principle, we merely need to ``invert" the vector field to map one image onto its predecessor. Using this approach for the kinds of images of interest here imposes stringent accuracy requirements on the model. Consequently, most publicly available optical flow models, which are typically designed for and trained on everyday images, fail to meet these precision demands. \FM uses a trainable registration model that can be adapted to the specific motion characteristics of each imaging modality. Drawing inspiration from synthetic datasets (e.g., Kitti~\cite{menze2015object}, Flying Chair~\cite{dosovitskiy2015flownet}), which leverage computer graphics techniques to generate optical flow fields between paired images, we developed an automated synthetic training data generation method that mimics the motion of an image acquisition device.  This emulation of controlled camera motion makes it possible to extract video frames and optical flow pairs to then use in training to fine-tune an optical flow estimator, eliminating the need for labor-intensive pixel-level optical flow field annotations. To illustrate the importance of \FMs' fine-tuning step we show results using the untuned DNN in \FM on examples from all 7 modalities in Fig.~\ref{fig:OffShelf}

\FM, by employing deep learning, inherently benefits from larger and more diverse training datasets, much like other DNN approaches. However, acquiring extensive medical imaging datasets poses significant challenges due to patient privacy considerations, the high cost of specialized equipment, and the labor-intensive nature of data acquisition. Our automated training data generation strategy addresses this problem, but the dataset sizes attainable by \FM remain substantially smaller compared to typical deep-learning benchmarks, particularly when only raw videos are available. When real mosaics from the same modality are available, larger synthetic datasets can be efficiently generated because these mosaics typically cover substantially larger tissue areas than single videos, thus providing richer data for training. Users should be aware, however, that, models fine-tuned on limited datasets may exhibit biases, necessitating the incorporation of inductive biases---domain-specific prior knowledge---into both data generation and model training.

Extending \FM to new imaging modalities depends critically on generating synthetic training data that closely resembles real data in terms of both image characteristics and optical flow patterns associated with camera motion. Our synthetic training methodology reliably captures general image features; however, additional modality-specific post-processing is typically required. For instance, circular masking replicates dark peripheral regions characteristic of fetoscopy, while carefully emulating illumination conditions can significantly improve realism for modalities with non-uniform lighting.

A particularly challenging aspect of synthetic data generation involves accurately simulating camera motion and non-rigid deformation, as this strongly influences registration performance. The level of non-rigid deformation in simulations must reflect realistic scenarios specific to the target modality. For example, modalities involving substantial parallax (e.g., laparoscopy, endoscopy) or direct tissue contact (e.g., reflectance confocal microscopy (RCM)) tend to experience pronounced non-rigid deformations, whereas stable setups (e.g., open-top light-sheet microscopy (OTLS)) typically exhibit smaller deformations. Selecting suitable distortion parameters is thus crucial and ideally should be informed by prior domain knowledge.

However, when such prior knowledge is limited or unavailable, a practical strategy involves generating an initial small dataset with varying deformation levels and qualitatively evaluating these against actual modality-specific data. This preliminary evaluation typically requires examining only a small set of synthetic examples to determine the extent of distortion introduced by different parameter settings. In contrast, when only mosaic images are available, increasing the size and diversity of the synthetic dataset helps mitigate the uncertainty arising from the absence of direct temporal reference. Although iterative validation under these circumstances can be challenging, our empirical results confirm that adopting this strategy successfully yields effective synthetic training datasets, substantially improving the accuracy and robustness of \FM across diverse imaging scenarios.

Finally, another practical consideration for achieving \textit{effective real-time deployment} of \FM is video frame rate management. Empirically, we observed robust optical flow estimation with at least 50\% overlap between consecutive frames. If high-frame-rate input videos have greater overlap,  selective downsampling can substantially accelerate processing without compromising accuracy.

While reliance on simulated data allows us to automatically tune our optical flow algorithm, it is also a major limitation of the current implementation of \FMs. In particular, in the current study, we concentrated on close-proximity modalities. Clearly, there is a need for further testing in more general imaging modalities where the imaging device is farther from the tissue. Such modalities may introduce additional distortions that are not accounted for by our current data simulation model. To ensure robust clinical applicability, future studies should ideally test \FM on real-world clinical data and fine-tune the model as necessary, including the distortion parameters used to generate training data, to adapt to unique characteristics of a particular clinical or laboratory setting.

Another limitation of the current \FM pipeline is its necessity for a GPU for real-time processing. While \FM runs well on affordable commercial GPUs, most clinical imaging systems do not meet this requirement. One potential solution is to replace the existing optical flow model with one that operates efficiently on CPUs or to use model compression techniques like pruning, quantization, or knowledge distillation~\cite{han2015deep,gou2021knowledge}. However, our DNN model's flexibility and GPU implementation offer robust and fast performance, leaving the trade-offs of using a CPU or non-DNN model uncertain.

One promising application of \FM is in intraoperative imaging systems, where real-time feedback can enhance surgical guidance. While our early studies indicate feasibility~\cite{yelamos2017margin,flores2019peri}, further research is needed to evaluate \FM's full potential and limitations in real-time settings. Future work should assess both the technical robustness of \FM under clinical conditions and its influence on surgical outcomes and workflow efficiency.

We plan to explore the clinical use of \FM in oral imaging, which currently relies on a telescopic probe with a limited field of view due to space constraints. Our previous studies on videomosaicking~\cite{Kose_Scientific,peterson2019feasibility} have shown its potential for increased imaging coverage. Integrating FM in this setting will provide real-time feedback to clinicians, enhancing imaging efficiency and diagnostic capabilities, ultimately improving patient outcomes.

\FloatBarrier

\bibliography{MSKInformatics}

\end{document}